# WEB INTERFACE FOR REFLECTIVITY FITTING

M. Doucet[a]*, R. M. Ferraz Leal[a], T. C. Hobson[a]

[a] Neutron Scattering Division, Oak Ridge National Laboratory, 1 Bethel Valley Rd., Oak Ridge, TN 37831, USA

*Correspondence e-mail: doucetm@ornl.gov

**Abstract** The Liquids Reflectometer at Oak Ridge National Laboratory's Spallation Neutron Source provides neutron reflectivity capability for an average of about 30 experiments each year. In recent years, there has been a large effort to streamline the data processing and analysis for the instrument. While much of the data reduction can be automated, data analysis remains something that needs to be done by scientists. For this purpose, we present a reflectivity fitting web interface that captures the process of setting up and executing fits while reducing the need for installing software or writing Python scripts.

## 1. Introduction

The Liquids Reflectometer (LR) is one of the two reflectometers at the Oak Ridge National Laboratory's Spallation Neutron Source (SNS) (Mason *et al.*, 2006). Specular reflectivity allows us to probe the depth profile of thin planar films by measuring the reflected distribution of beams of neutrons or X-ray photons scattered off their surface. In the case of the LR, neutron reflectivity measurements allow us to probe layer structures with length scales between a few Angstroms and a few thousand Angstroms.

The reflectivity distribution $R(Q)$ is measured as a function of momentum transfer $Q=4\pi \sin(\theta)/\lambda$, where $\theta$ is the scattering angle and $\lambda$ is the neutron wavelength. $R(Q)$ depends on the scattering length density (SLD) distribution along the axis perpendicular to scattering plane and is approximately given by (Sivia, 2011):

$$R(Q) \approx \frac{16\pi}{Q^4} \left| \int_{-\infty}^{\infty} \frac{d\beta}{dz} e^{-izQ} dz \right|^2$$

For a given compound, the scattering length density $\beta$ is related to the scattering length of each atom in the system:

$$\beta = \rho N_A / m \sum_{i=1}^{n} b_i$$

where $\rho$ is the mass density of the compound, $m$ is its mass, $N_A$ is Avogadro's number, and $b_i$ is the coherent scattering length of the $i^{th}$ atom in the system. Fitting the reflectivity distribution thus gives us the SLD profile of our sample as a function of depth. This profile in turn gives us information about the chemical composition of our layered system.

The calculation of the reflectivity profile from a layered system is generally done by following the reflected and transmitted particle waves at the layer boundaries (Abeles, 1948; Parratt, 1954). This approach gives an accurate reflectivity value in all $Q$ ranges, as opposed to the $R(Q)$ equation above which holds for angles greater than the angle of total reflection.

The LR provides data for an average of about 30 experiments each year as part of the SNS User Program, producing about 10000 data sets per year. Each experiment typically takes three to five days, with some taking only a single day and some taking up to a week. It is important to automate and simplify data treatment as much as possible. Once acquired, the LR data is automatically reduced using the Mantid framework for data analysis and visualization (Arnold *et al.*, 2014). In most cases, users are able to leave the laboratory with reduced $R(Q)$ reflectivity data ready to be analysed. Subsequent modelling of reflectivity data is usually done using Motofit (Nelson, 2006) or REFL1D (Kienzle *et al.*). Other software packages such as GenX (Björck & Andersson, 2007) or custom user software are also used. Motofit is a package that provides reflectivity modelling within the IGOR Pro environment [i]. In addition to its minimization and error analysis capabilities, Motofit provides a graphical user interface. REFL1D is a Python package originally developed by the DANSE Project [ii] that also provides minimization and error analysis of reflectivity models. It does not provide a graphical user interface.

---

[i] https://www.wavemetrics.com/

The choice of reflectivity fitting software is generally done with the help of SNS staff. Instrument scientists also train users in the use of the chosen software. For this reason, developing tools to simplify that process and empower visiting scientists to fit their data has been a focus of the LR staff for the past few years.

In the present article, we describe a web application developed as a front end to REFL1D. It integrates with the existing web monitoring site available at the SNS (Shipman *et al.*, 2014) and allows users to visualise and fit their reflectivity data using a simple set of web forms. It allows an easy transition from modelling activities done during their experiment and analysis work done at their own institution. The application code is available at https://github.com/neutrons/web_reflectivity.

## 2. Application description

### 2.1. Implementation

The application is designed to let users define a model for a given data set using a web form to set up their layer structure and fit parameters. Once the model is submitted, a fit job is translated into a Python script to perform the minimization using REFL1D. Information related to the model and fit status is kept up to date in the application database, which allows users to come back to a particular data set, view fit results, modify their model and resubmit as necessary. The web application was developed in Python using the Django web framework [iii]. It is hosted on a Red Hat Enterprise Linux node running Apache with mod_wsgi [iv] and uses a PostgreSQL database [v].

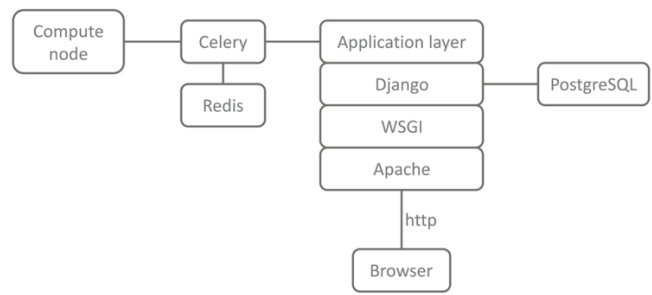

**Figure 1** Overview of the reflectivity fitting web interface, showing the software packages used to build the application.

Fits are performed on one of the SNS analysis nodes for the user. This allows us to separate the resources needed to host the application from the resources needed to perform calculations, making the application more easily scalable and allowing for the use of a different compute resource architecture in the future should the need arise.

The web application is available to all users with a facility account. The web application authenticates facility users and ensures that users only have access to their own data. Every piece of model data stored in the application database is tied to a particular user.

Job management is done using a Django remote submission package developed in-house (Hobson *et al.*, 2017). It manages remote jobs using the Celery distributed task queue [vi] and provides real-time monitoring of remote jobs and their associated logs. Celery uses message brokers to pass messages between the Django application and compute nodes. The Redis [vii] in-memory data structure store is used as the message broker. The script executed on the compute node sets up and executes the REFL1D fit, then gathers the output data.

Interactive plots are generated using Plotly [viii] for Python. Plotly produces graphics of various types with a rich set of options. In addition, the plots generated by Plotly can be pushed to the Plotly web site for further customization and production of publication quality graphics.

---

[ii] http://danse.us/
[iii] https://www.djangoproject.com/
[iv] https://github.com/GrahamDumpleton/mod_wsgi/
[v] https://www.postgresql.org/

[vi] http://www.celeryproject.org/
[vii] https://redis.io/
[viii] https://plot.ly/python/

## 2.2. Features

### 2.2.1. Integration with existing web tools

The web application for reflectivity fitting was designed to integrate with existing SNS web applications. Facility users are already able to follow the progress of their experiment using a web monitor (Shipman *et al.*, 2014). In addition to reporting the current status of each instrument, the web monitor allows users to browse through the reduced reflectivity data for their experiment. The web monitor is then used as the entry point of the reflectivity fitting application.

### 2.2.2. Available engines

The reflectivity application can generate REFL1D scripts that use three fitting engines: a Levenberg-Marquardt algorithm (Levenberg, 1944; Marquardt, 1963), a Nelder-Mead algorithm (Nelder & Mead, 1965), and the DREAM algorithm (Vrugt *et al.*, 2009). The DREAM algorithm is a hybrid algorithm that uses differential evolution to adapt the evolution of Markov chains. REFL1D uses the DREAM algorithm to infer the probability distribution of fit parameters and provide an error analysis that gives us uncertainty estimates for each fit parameter and a correlation plot between those parameters. The uncertainties are reported on the fit result page of the web application. The additional files produced by REFL1D, which include the correlation plot, theoretical SLD profiles, and theoretical reflectivity profiles, are saved in the user's home directory on the SNS file system.

### 2.2.3. Functional constraints

One strength of REFL1D is the ability to process complex constraints written as a function of fit parameters. The web interface allows users to add multiple constraints between the layer parameters of a model. Equations written in Python can be entered. A dedicated page is available to validate and manage the constraints for a given model.

### 2.2.4. Simultaneous fitting

The web application offers an interface to set up the simultaneous fit of multiple data sets. The results of those simultaneous fits are stored separately in the application database so that several fits involving the same data sets can be done without interfering with each other. Users can tie model parameters from data sets selected to be co-refined by simple drag-and-drop. Once those constraints are set and the fit is executed, the reflectivity and SLD plots of the resulting fits are presented along with the output parameters.

### 2.2.5. File handling

The web application allows users to upload their own reflectivity data. This data is stored in a database and is tied to the user rather than the experiment. Data from other x-ray and neutron facilities can be uploaded to be co-refined with LR data. A data manager page is available for users to annotate and sort their data.

### 2.2.6. Model handling and sharing

Several features are available to handle models. Fit results can be saved independently in order to re-use models for different data sets without having to re-enter them by hand. A model list is provided to annotate and manage models. Users can pull models from their list when fitting a new data set. Since members of a research team may want to work together on analysing their data, it is also possible to share models between users by sharing a URL that contains an encoded identifier for the model to be shared.

Depending on the experiment, a sample can be measured by reflecting the neutron beam on either side of the film. For this purpose, a link is provided to reverse a model. For instance, a measurement done with the neutrons impinging directly on the top layer surface may be followed by a measurement with the neutrons coming through the substrate. When building a layer model for the second case, one has to reverse the order of the layers so that the front layer exposed to the incoming neutrons in the first case now becomes the back layer. When reversing the order of the layers, special care has to be taken to making sure that the roughness parameter between two layers is assigned to the right layer. The model reversing feature keeps track of the roughness values and simplifies the process of using a model in either geometry.

## 3. Usage examples

The web application is best demonstrated with examples using LR data. In this section, we describe an example of a simple fit, and the co-refinement of two related data sets.

## 3.1. Single reflectivity model

The following example is a reflectivity measurement made on a thin film (see Figure 2) with layers of copper and amorphous silicon deposited on a silicon substrate, topped by a spin-coated layer of polyacrilic acid (PAA). When modelling the structure of the film, we also allow for a native oxide layer on the surface of the silicon substrate. The reflectivity measurement was done at the LR with the incident beam coming through air and impinging on the PAA surface of the film. This data was automatically reduced by the SNS post-processing system so that it was immediately ready to fit. The model form for this data set is shown on Figure 3. For each selected model parameter, a value range can be entered in the bottom section of the form. The overall $Q$ range used for the fit can also be modified.

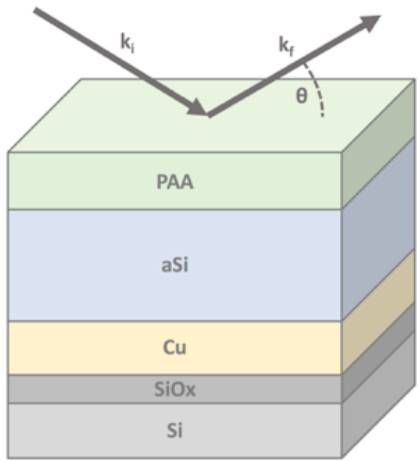

**Figure 2** Thin film structure for the example data set, consisting of a deposited layer of Cu on a silicon substrate, a deposited layer of amorphous silicon, and a spin-coated layer of PAA. SiOx denotes the silicon substrate's native oxide.

The fit page notifies the user once the REFL1D fit is complete. Once the fit page is refreshed, the parameter values are updated in the model table and in the fit parameter range table, as shown on Figure 3. The $R(Q)$ data plot is updated with the output model, and the SLD profile of this model is also shown. If the DREAM algorithm is used, the uncertainty on each fit parameter is displayed. Results from several fit jobs can be overlaid for comparison.

Varying parameters can be selected with using the check box next to each parameter. Once fit results are available, the uncertainty on each fit parameter is displayed in the "Fitting parameters" section if the chosen minimization technique estimates it.

## 3.2. Co-refinement

Sometimes two measurements taken with the same sample in two configurations can help extract better information about the structure of a film. In the example above, we have tried to measure the thicknesses, SLDs, and roughness values of our native oxide, copper, amorphous silicon, and PAA. To constrain the fit, we can also use a measurement done on the film before spin coating the PAA. Since depositing the PAA is not expected to change the rest of the film, we can impose that the unmodified layers have the same parameters in both models. We can fit the two data sets in a single minimization problem and tie together parameters that should remain the same before and after the PAA deposition. If we allow for the surface roughness parameter of the amorphous silicon to change after depositing of PAA, we are left with nine constraints between the two models. Using the simultaneous fitting form shown in Figure 4, such constraints can be entered by dragging and dropping parameters between the two models.

Once constraints between models have been set, clicking the "perform fit" button submits the job for execution. The starting parameters for the models will be those from the existing individual fits. Once the simultaneous fit is completed, users can see their fit results on the simultaneous fit page. The fit results for $R(Q)$ are plotted, and the model table of Figure 4 now shows the output values of the fit parameters. Uncertainties are also shown if they are estimated by the minimization.

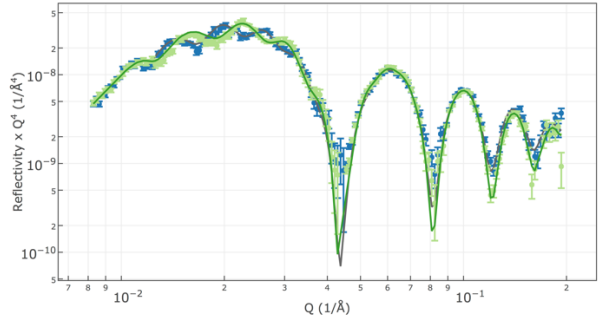

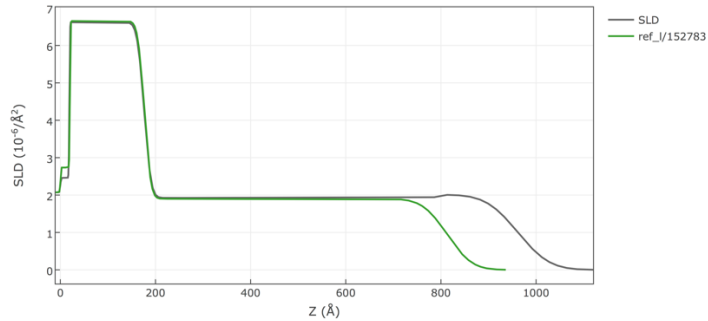

**Figure 3** Model fitting form for our example thin film with PAA (identified as run ref_l/152922 on this figure). The same film measured before the PAA deposition (labelled ref_l/152783) is also plotted for comparison.

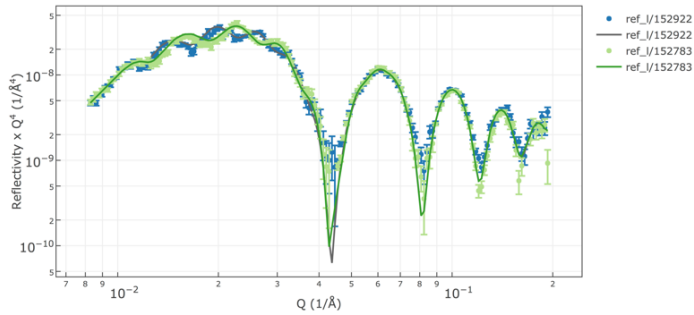
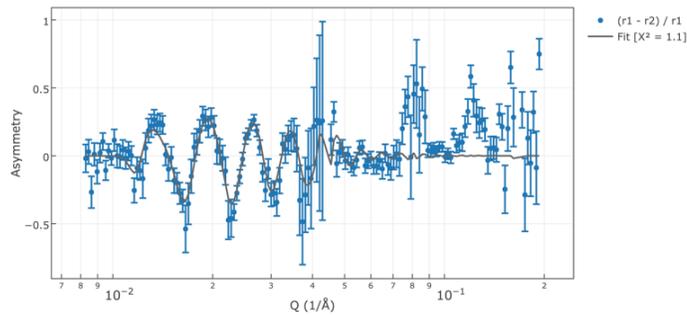
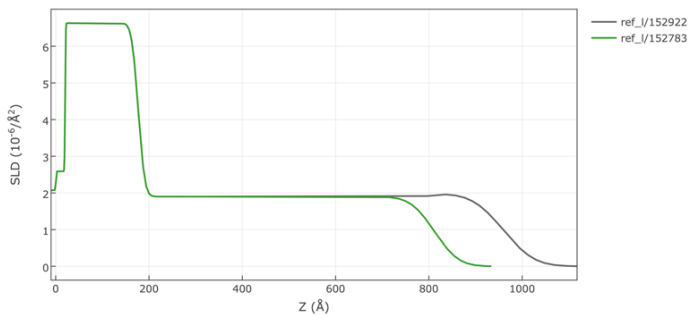

**Figure 4** The simultaneous fit page, showing the data and models of two data sets. The constraint matrix for the simultaneous fit job is shown at the bottom of the page. Constraints are added by dragging and dropping parameters between the two models. Tied parameters are indicated by a blue label. Clicking this label will remove the constraint and revert the parameter to its original value.

## 4. Summary

The number of experiments performed at the Liquids Reflectometer at the SNS demands a streamlining of data processing so that users leave the laboratory with quality data sets and a clear plan for how they are going to analysis them. An important part of this effort is to offer analysis tools that empower users to work independently regardless of whether they have a background in scientific computing or not. To address this need, we developed a user interface that simplifies the work of modelling reflectivity data. By deploying it as a web application, we have also greatly reduced the problem of installing software on various platforms, and we have made the transition from the laboratory to the home institution seamless. The reflectivity fitting application lets users manage their fit jobs, their models, and their data files. It provides a simple interface that lets users concentrate on the science by removing the need to install software or write Python scripts.

**Acknowledgements** This research was sponsored by the Division of Scientific User Facilities, Office of Basic Energy Sciences, US Department of Energy, under Contract no. DE-AC05-00OR22725 with UT-Battelle, LLC. A portion of this research used data from the Liquids Reflectometer at the Spallation Neutron Source. The REFL1D package used in this work was originally developed as part of the DANSE Project under NSF award DMR-0520547. Special thanks to Jim Browning and John Ankner for immediately starting to use the interface and giving us feedback. MD would like to thank collaborators Jim Browning, Josh Kim, Katie Browning and Gabe Veith for allowing us to use the data shown in this article. Finally, MD would also like to thank Paul Kienzle for developing and maintaining REFL1D, and for all the discussions on the topic of software and minimization. This manuscript has been authored by UT-Battelle, LLC under Contract No. DE-AC05-00OR22725 with the U.S. Department of Energy. The United States Government retains and the publisher, by accepting the article for publication, acknowledges that the United States Government retains a non-exclusive, paid-up, irrevocable, world-wide license to publish or reproduce the published form of this manuscript, or allow others to do so, for United States Government purposes. The Department of Energy will provide public access to these results of federally sponsored research in accordance with the DOE Public Access Plan (http://energy.gov/downloads/doe-public-access-plan).